\documentclass[preprint,showpacs,preprintnumbers,amsmath,amssymb]{revtex4}
\usepackage{graphicx}
\usepackage{dcolumn} 
\usepackage{bm}      
\begin{document}
\title {Effect of a bunch shape on its TMCI spectrum and threshold with high space charge}
\author{V. Balbekov}
\affiliation {Fermi National Accelerator Laboratory\\
P.O. Box 500, Batavia, Illinois 60510}
\email{balbekov@fnal.gov} 
\date{\today}

\begin{abstract}

Transverse mode coupling instability of bunched beam is investigated in the paper at different form of the bunches with space charge included. 
Equation of transverse motion of the bunch in parabolic potential well of synchrotron oscillations is derived and analysed.
The bunch of constant density (flat bunch) is examined in detail to make comparison with the square well model.
It is shown that both models result in very close instability thresholds of the flat bunch.
Then different form bunches are investigated in the parabolic potential well.
It is shown that decrease of the bunch r.m.s length leads to the growth of its threshold, that is the flat bunch model gives only a minimal estimation of the threshold. 
The results are treated in terms of Landau damping due to spread of the space charge tune shift.

\end{abstract}
\pacs{29.27.Bd} 
\maketitle

\section{INTRODUCTION}
\vspace{-5mm}
Transverse mode coupling instability (TMCI) of the bunched beam has been observed first in the electron storage ring PETRA and explained by Kohaupt \cite{Koh}.
Using the two-particle model, the author have shown that the instability occurs when tunes of two head-tail modes approach each other being shifted by the bunch wake field.  
The shift should be about synchrotron tune to reach the tune coalescence and the beam instability.

In proton rings, the space charge (SC) tune shift $\Delta Q_{sc}$ has to be taken into account as well because it is typically exceeds the synchrotron tune $Q_s$.
This effect has been considered first by Blaskiewicz \cite{Bl1,Bl2}. 
The main point of the papers is that the SC pushes upward the TMCI threshold that is improves the beam stability.
However, relatively small SC tune shift was considered in these works.

Different wakes at unlimited value of the SC tune shift were considered in the subsequent papers  [4-9].
It was shown that, at increasing tune shift, the TMCI threshold goes up if the wake is negative, and goes down if it is positive. 
The effect of oscillating wake is more complicated being dependent on its phase advance in the bunch \cite{Bu3,Ba3}.

Another result has been represented in Ref.~\cite{Bl3}.
According to it, the stable and unstable zones alternately replace each other when the SC tune shift increases, even at the constant wake field.
The two-particle approximation has been used in the paper like the pioneer work \cite{Koh}, with a model of the space charge field proposed. 
Further examination of the model looks to be needed.

It should be noted that most of the mentioned results were obtained in framework of the flat (square) bunch model.
It ignores dependence of the SC tune shift of particles on their longitudinal position, and related betatron tune spread. 
Meanwhile, it is known that the spread due to nonlinearity or chromaticity affects the transverse instability thresholds by a redistribution of the tunes \cite{L1}. 
Similar effect should not be excluded if the spread is caused by space charge. 
This problem is investigated in this paper.

In Sec. II, the physical model is introduced, and equation of transverse oscillations of arbitrary bunch with wake field and space charge is proposed.

In Sec. III, the equation is applied to the flat bunch to compare the results with the known models and to check their identity and applicability. 

In Sec. IV, the equation is applied to different bunch forms to investigate their spectra and thresholds in comparison with the flat bunch.
 
\section{PHYSICAL MODEL}
\subsection{General}

We will consider coherent transverse oscillations of a bunch in the rest frame, representing its displacement in a point of the longitudinal phase space as real part of the function
\begin{eqnarray}
X = Y(\theta,u)\exp\big[-i(Q_0+\xi)\theta-i(Q_0+\nu)\Omega_0t\big] 
\end{eqnarray}
where $\theta=z/R_0$ and $u=\dot\theta$ are the longitudinal coordinate and corresponding velocity, $R_0$ and $\Omega_0$ are the accelerator average radius and central revolution frequency, $Q_0$ and $\nu$ are the bare betatron tune and the addition to it caused by the collective effects, and $\xi$ is the normalized chromaticity:
\begin{eqnarray}
\xi=-\frac{dQ_0/d\ln(u)}{\alpha-1/\gamma^2}
\end{eqnarray}
with $\alpha$ as the momentum compaction factor.
The function $Y$ satisfies the equation \cite{Ba1}
\begin{eqnarray}
\nu Y+i Q_s\frac{\partial Y}{\partial\phi}+\Delta Q_{sc} (Y-\bar Y) =
2\int_\theta^\infty q(\theta'-\theta)\exp [-i\xi(\theta'-\theta)] 
\bar Y(\theta')\rho(\theta')\,d\theta'
\end{eqnarray}
where $\phi$ and $Q_s$ are phase and tune of the synchrotron oscillations, 
$\Delta Q_{sc}(\theta)$ is the space charge tune shift, and  $q(\theta)$ is the normalized wake field potential which is proportional to usual transverse wake function $W_1(z)$:
\begin{eqnarray}
q(\theta) = \frac{r_0 N_b R_0W_1(-R_0\theta)} {8\pi\beta^2\gamma Q_0}
\end{eqnarray}
with $r_0=e^2/mc^2$ as the particle electromagnetic radius, $\beta$ and $\gamma$ as its normalized velocity and energy, and $N_b$ as the bunch population.
Besides, the notation is used in Eq.~(3):
\begin{subequations}
\begin{align}
\int_{-\infty}^\infty F(\theta,u)Y(\theta,u)\,du = \rho(\theta)\bar Y(\theta),  \\
\int_{-\infty}^\infty F(\theta,u)\,du  = \rho(\theta),                   \qquad \\
\int_{-\infty}^\infty \rho(\theta)\,d\theta = 1                     \qquad\quad\;\,
\end{align}
\end{subequations}
where $F$ is the bunch distribution function in the longitudinal phase space.

\subsection{Used simplifications}

Being interested mainly by dependence on the TMCI threshold on the SC tune shift at arbitrary bunch shape, we restrict ourselves to the case of zero chromaticity and constant wake: $\xi=0,\;q=q_0=const$. Besides, we will consider only linear synchrotron oscillations taking $Q_s=1$ in the equations. 
Finally, we will consider a bunch of restricted length, and will take  new longitudinal coordinate and velocity $x\propto\theta,\;v=\dot x$, to have the bunch location in the region $-1<x<1$.
Then Eq.~(3) obtains the form
\begin{eqnarray}
\nu Y+i\frac{\partial Y}{\partial\phi}+\Delta Q_{sc} (Y-\bar Y) = 2q_0\int_x^1 \bar Y(x')\rho(x')\,dx'
\end{eqnarray}
with the normalization conditions replacing Eq.~(5)
\begin{subequations}
\begin{eqnarray}
\int F(x^2+v^2)Y(x,v)\,dv = \rho(x)\bar Y(x),\\
\int F(x^2+v^2)\,dv = \rho(x), \qquad\, \\
\int_{-1}^1\rho(x)\,dx = 1.\qquad\quad\;\;
\end{eqnarray}
\end{subequations}
Parameter $q_0$ is proportional to an average value of the original wake potential $W_1$, if the last is a monotonous function of the coordinate (for example the resistive wall wake).
The proportionality coefficient is clear from Eq.~(6) itself because it should give the tune shift $\nu=q_0$ for the lowest head-tail mode $Y=\bar Y=1$:
\begin{eqnarray}
\nu=2\int_{-1}^1\rho(x)\,dx\int_x^1q(x-x')\rho(x')\,dx' 
\rightarrow 2q_0\int_{-1}^1\rho(x)\,dx\int_x^1\rho(x')\,dx'=q_0.\nonumber
\end{eqnarray}

\subsection{High space charge approximation}

For further, it is convenient to consider only the even part of the function $Y$ which has the form  $\;Y^+(\phi)=[Y(\phi)+Y(-\phi)]/2$ and satisfies the second order equation \cite{Ba3}:
\begin{eqnarray}
\frac{\partial}{\partial\phi}\left(\frac{\partial Y^+}{\hat\nu \partial\phi}\right) 
+\hat\nu Y^+ =\Delta Q_{sc}\bar Y+2q_0\!\!\int_x^1\!\!\bar Y(x')\rho(x') dx'
\end{eqnarray}
where $\hat\nu(x)=\nu+\Delta Q_{sc}(x)$.
Multiplying this equation by the function $F$, integrating on $v$, using Eq.~(7), and taking into account the relations
\begin{eqnarray}
\frac{\partial}{\partial\phi}=v\frac{\partial}{\partial x}
-x\frac{\partial}{\partial v},\qquad \frac{\partial F}{\partial\phi}=0
\end{eqnarray}
obtain
\begin{eqnarray}
\frac{d}{dx}\left(\frac{d}{\hat\nu dx}\int Y^+Fv^2\,dv\right)
+\frac{d}{dx}\left(\frac{x\rho\bar Y}{\hat\nu}\right)+\nu\rho\bar Y 
= 2q_0\rho(x)\int_x^1\bar Y(x')\rho(x')\,dx'
\end{eqnarray}

Similar equation has been proposed earlier in framework of the square potential well model, where the positive phase can be identified with the longitudinal coordinate \cite{Ba2}.  
The goal of this paper is to extend this to arbitrary potential well using the high space charge approximation developed in papers \cite{Bu1,Ba4}. 
The main idea is that, at rather large $\Delta Q_{sc}$, the function $Y^+$ depends  on coordinate $x$ essentially stronger then on velocity $v$.
In accordance with this, one can put $Y^+\simeq\bar Y$ in the left hand part of Eq.~(10).
It results in the equation for the bunch dipole momentum 
$D(x)=\rho(x)\bar Y(x)$:
\begin{eqnarray}
\frac{d}{dx}\left(\frac {d(U^2 D)}{\hat\nu dx}\right) 
+\frac{d}{dx}\left(\frac{xD}{\hat\nu}\right)+\nu D =
2q_0\rho(x)\int_x^1 D(x')\,dx'
\end{eqnarray}
where
\begin{eqnarray}
\rho(x)U^2(x)=\int_{-1}^1 F(x^2+v^2)\,v^2\,dv
\end{eqnarray}
It is easy to check that
\begin{eqnarray}
\frac{d(\rho U^2)}{dx}=-x\rho(x).
\end{eqnarray}
A usage of this relation in Eq.~(11) would allow to get a final form of this second order 
equation with the integral.
However, it is more convenient for the numerical solution to represent it as the system of the first order equations
\begin{subequations}
\begin{align}
\frac{dD}{dx}\, = \,\frac{\hat\nu(x)D_1(x)}{U^2(x)},\qquad\qquad\qquad\qquad\qquad\\
\frac{dD_1}{dx}=\left[\frac{d}{dx}\left(\frac{U^2\rho'(x)}{\hat\nu\rho}\right)-\nu\right]D + 2q_0\rho D_2, \\ 
\frac{dD_2}{dx}=-D(x).\qquad\qquad\qquad\qquad\qquad\qquad
\end{align}
\end{subequations}
The incoming functions satisfy the relation $\rho(x)=0$ and $U^2(x)=0$ at $x=\pm 1$.
Because the bunch dipole momentum $D(x)$ and its derivative are expected to be constrained values in any point, following boundary conditions must be satisfied:
\begin{eqnarray}
 D_1(\pm 1) = 0, \qquad D_2(1) = 0.
\end{eqnarray}
Therefore numerical solution of the problem can be performed using 
the following steps, as it has been proposed and applied in Ref.\cite{Ba2,Ba3}:

1. At given value of the SC tune shift, and with a trial value of the wake $q_0$, series of Eq.~(14) should be  resolved step by step by moving from the bunch end $x=1$ to its beginning $x=-1$, with initial conditions $D(1)=1,\; D_1(1)=0,\; D_2(1)=0$.

2. The tune $\nu$ should be varied to find the value providing fulfillment of the boundary condition $D_1(-1)=0$.

3.  Steps 1-2 should be repeated  with different $q_0$ so much times to obtain several curves $\nu_i(q_0)$ describing dependence of eigenvalues on the wake strength at given 
$\Delta Q_{sc}$.

4. The instability threshold of each mode should be determined as the curve turning point.

The operations should be repeated with all desirable tune shifts.
Note that the condition $\hat\nu(x)\ne 0$ is assumed to be fulfilled at any step of the solving.
This important point will be discussed in detail in Sec.~IV.  



\section{Flat bunch}


The TMCI threshold of a flat bunch ($\rho=const$) was considered earlier using different approximations, including the expansion technique and/or the square well model \cite{Bl1,Bu1,Ba3,Bu2}. 
However, the case of lineal synchrotron oscillations has been analyzed only at a modest value of the SC tune shift.
Equation~(14) allows to consider the problem more widely.

The involved functions are at $|x|<1$
\begin{subequations}
\begin{align}
\rho(x) = \frac{1}{2},      \\
U^2(x)=\frac{1-x^2}{2},     \\
\hat\nu=\nu+\Delta Q_{sc}=const
\end{align}
\end{subequations}
Using the notations 
\begin{eqnarray}
{\cal P}=\hat\nu\nu=\hat\nu(\hat\nu-\Delta Q_{sc}),\qquad {\cal Q} = q_0\hat\nu
\end{eqnarray}
one can reduce Eq.~(11) to the form not including the SC tune shift explicitly:
\begin{eqnarray}
\frac{d}{dx}\left(U^2\frac{dD}{dx}\right)+{\cal P}D
={\cal Q} \int_x^1 D(x')\,dx'
\end{eqnarray}
Similar equation has been obtained in Ref.~[6] on the base of the square well model.
With an equalization of the bunch length, the only remaining difference would be the factor 
$4/\pi^2$ instead of $U^2(x)=(1-x^2)/2$ in Eq.~(18).

Solution of the equation by above described method provides an infinite set of eigentunes 
${\cal P}_n$ at each $\cal Q$. 
They form the lines in the $(\cal Q$--$\cal P)$ plane, some of them are shown in Fig.~1.
\begin{figure}[t!]
\includegraphics[width=90mm]{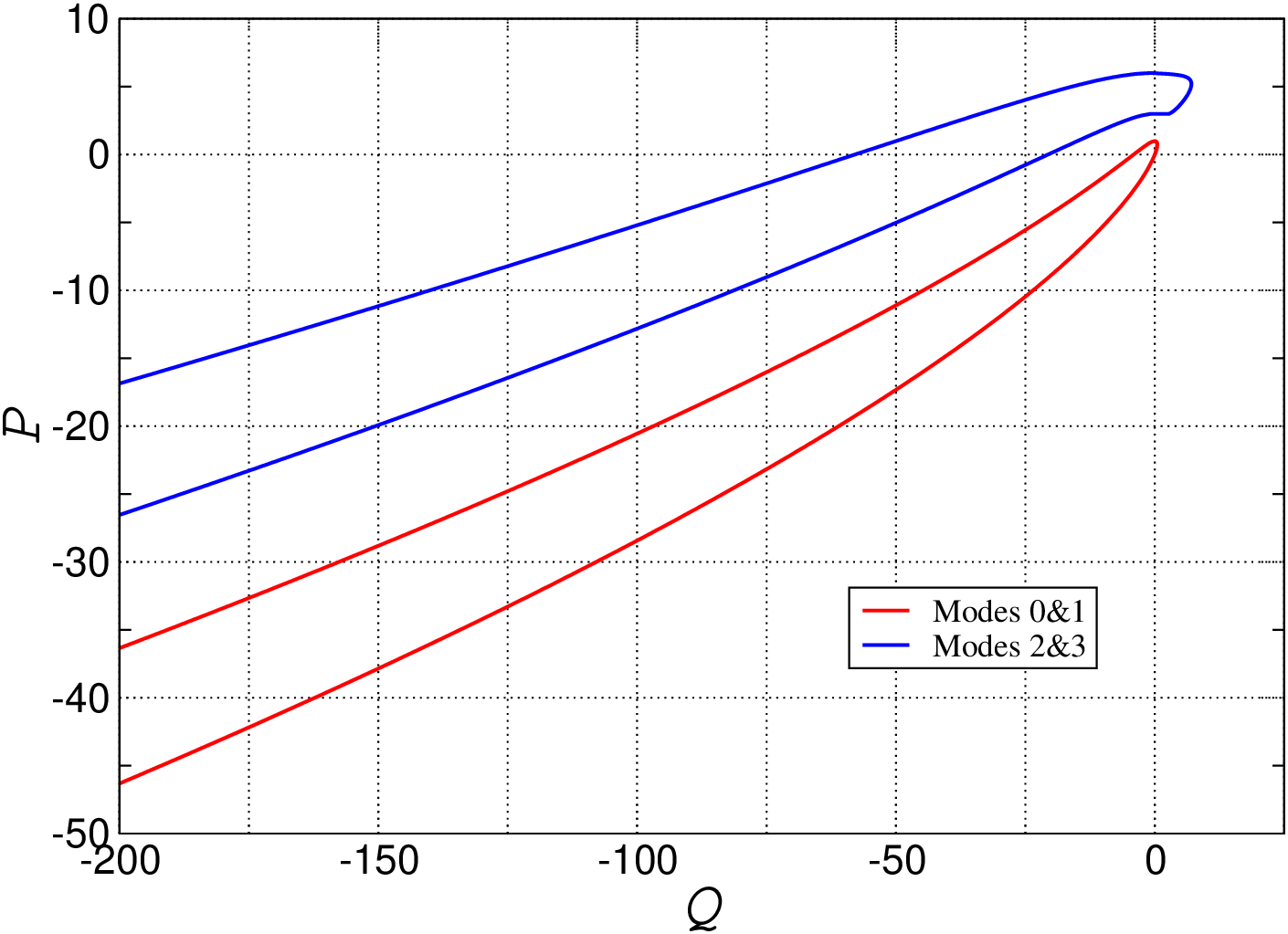}
\caption{Some eigentunes of Eq.~(18) against the referred wake strength.
The red line $M_{0,1}$ illustrates coalescence of the lowest head-tail pair,
the blue line $M_{2,3}$ does this for the next pair.} 
\end{figure}
Note that, at ${\cal Q}\!=\!0$, Eq.~(18) is Legendre equation with  eigennumbers 
${\cal P}_m=0,\,1,\,3,\,6,\,\dots\,m(m+1)/2,\,\dots$ .
The lines in Fig.~1 cross the vertical axis ${\cal Q}$ just in these points, so that the numbers $m$ can be treated as the indexes of the lower multipole in the solution. 
At ${\cal Q}\!>\!0$, some lines merge in pairs which will be marked further by symbols $M_{0,1},\, M_{2,3}$ etc.
For example, the red lines link up at ${\cal Q}\!=\!0.468$ producing the $M_{0,1}$ coupled mode.  
Actually, the lines extend to the region ${\cal Q}\!>\!0.468$ as the complex-conjugate pair, that is ${\cal Q}_{0,1}\!=\!0.468$ is the critical point of the $M_{0,1}$ mode.
Higher modes have higher critical points: for example ${\cal Q}_{2,3}\!=\!7.07$ (blue lines in Fig.~1), ${\cal Q}_{4,5}\!=\!28.6$,  etc. 

According to Eq.~(17), each curve of the $(\cal Q$--$\cal P)$ plane is mapped to 2 curves in the $(q_0$--$\hat\nu)$ plane, dependent on $\Delta Q_{sc}$:
\begin{eqnarray}
\hat\nu=\frac{\Delta Q_{sc}}{2}\pm\sqrt{\frac{\Delta Q_{sc}^2}{4}+\cal P},
\qquad q_0=\frac{\cal Q}{\hat\nu}
\end{eqnarray}
\begin{figure}[t!]
\begin{minipage}{0.49\linewidth} 
\includegraphics[width=75mm]{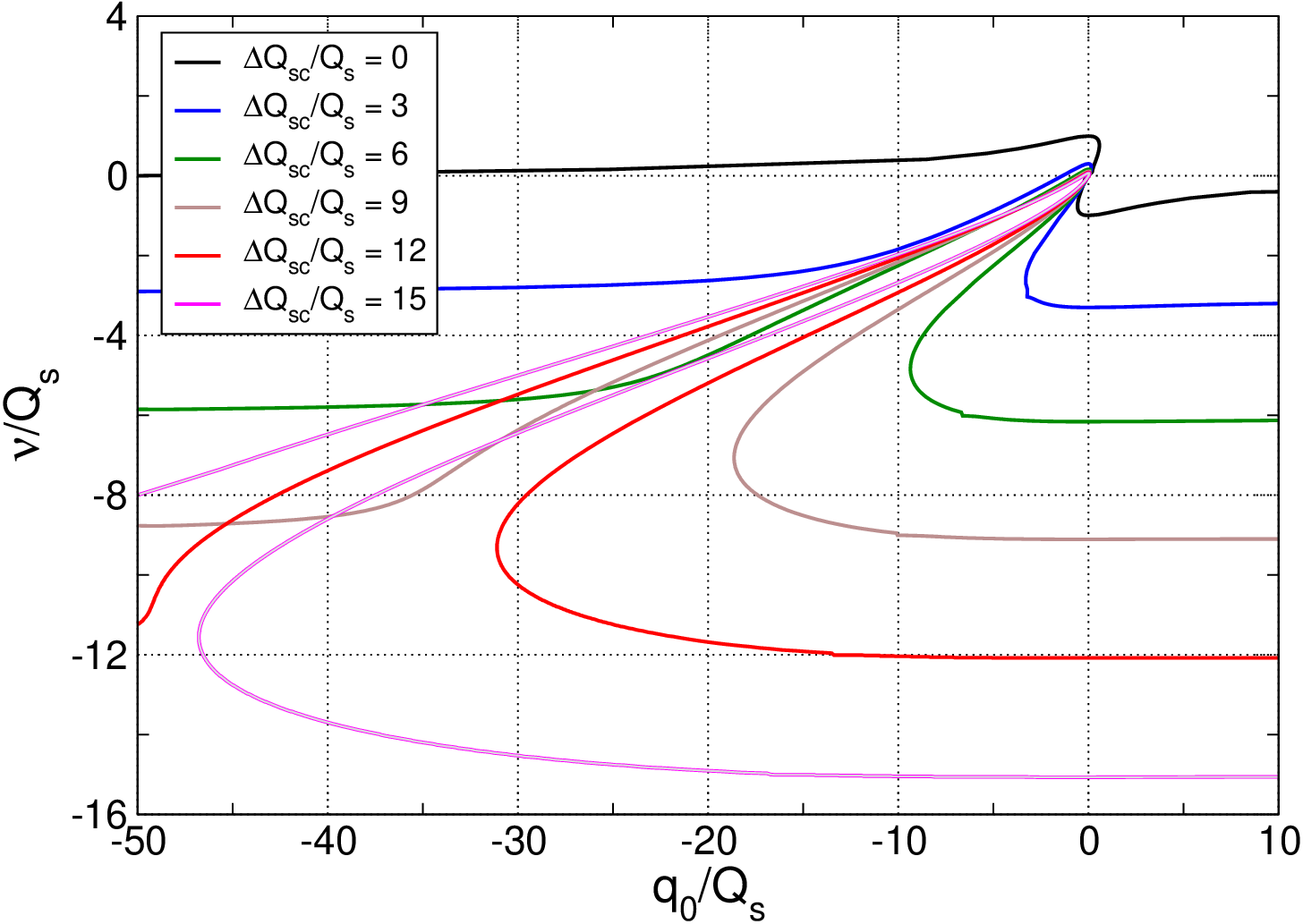}
\end{minipage}
\begin{minipage}{0.49\linewidth} 
\includegraphics[width=75mm]{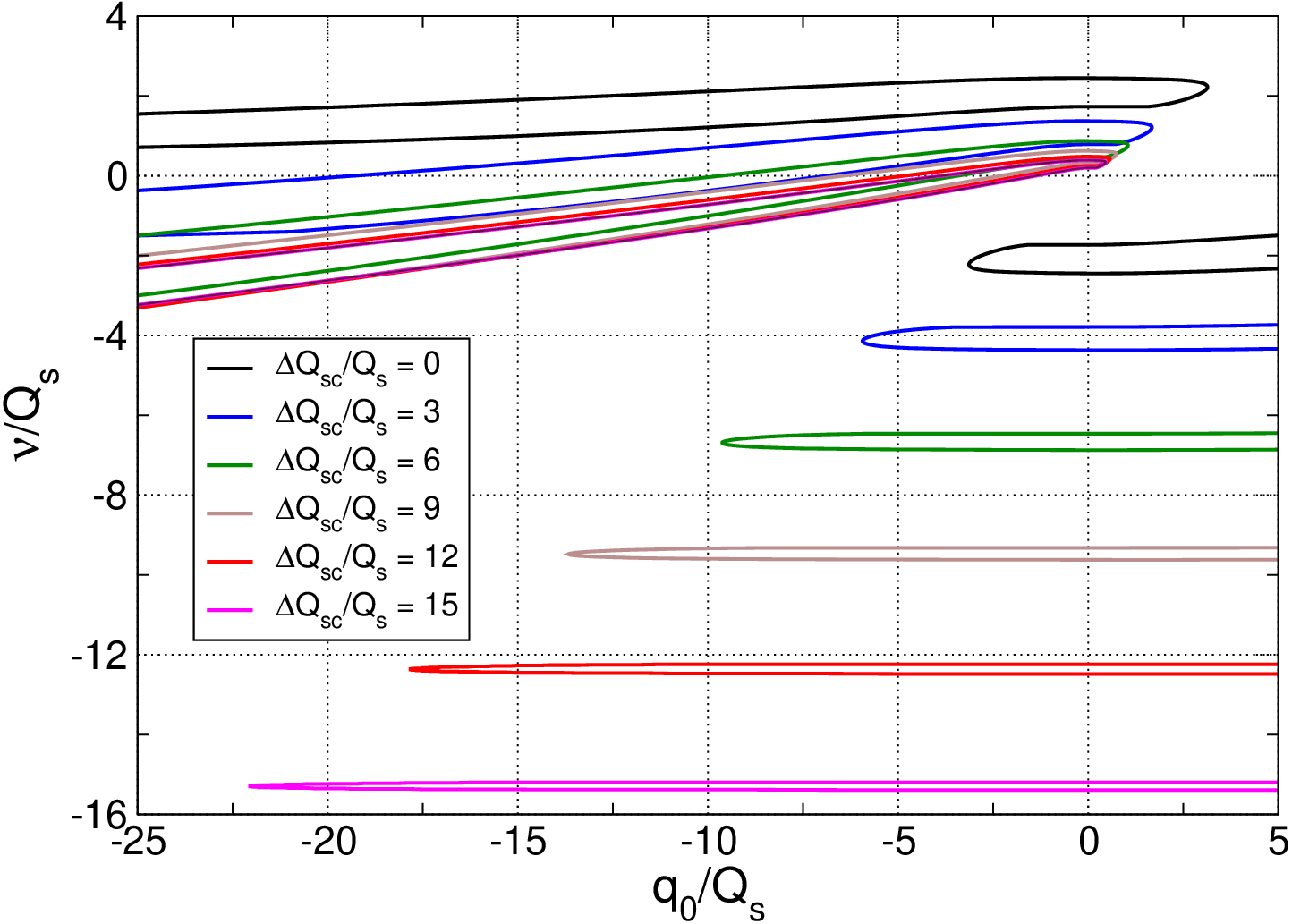}
\end{minipage}
\caption{Several eigentunes of the bunch against the wake strength at different value of the 
space charge tune shift. 
The lower bunch modes $M_{0,1}$ generated by red line of Fig.~1 are plotted in the top panel, 
and the modes $M_{2,3}$ are plotted in the bottom panel (blue line in Fig. 1).} 
\end{figure}
(see Fig.~2). Each obtained point is the eigentune of some head-tail mode with space charge and wake field.
There are left and right turning points in the curves where two real head-tail eigentunes join together.
It means appearance of the complex conjugated eigentunes that is threshold of corresponding TMCI mode.

For example, at $\,\Delta Q_{sc}\!=\!0,$  $M_{0,1}$ mode has the turning points 
$\,q_0/Q_s\!=\!\mp 0.57\,$ with  corresponding values of $\,\nu/Q_s\!=\!\mp 0.75$.
They are the TMCI thresholds of negative or positive wake without space charge (the black line in the upper panel of Fig.~2).
The curves stretch to the left-down direction when the SC tune shift increases, resulting in a movement of the turning points.
Change of the TMCI threshold in the process depends on sign of the wake: threshold of the
positive wake decreases going to 0 at $\Delta Q_{sc}\rightarrow\infty$ whereas threshold of the negative wake $|g_0|=-g_0$ tends to $\infty$ in the case.
Similar behavior of the $M_{2,3}$ mode is shown in the lower panel of Fig.~2.

The more important in practice case of negative wakes is additionally illustrated by Fig.~3 where the mentioned modes are shown together.
When the SC tune shift increases, both turning points move to the left but with different velocity.
Turning point of the lower mode $M_{0,1}$ moves especially quickly, particularly because of 
fast growth of distance between initial points of the loop in vertical axis, which always 
exceeds $\Delta Q_{sc}$.   
Other modes grow slower.
As a result, the mode $M_{0,1}$ is the most unstable only at $\Delta Q_{sc}/Q_s<6$, otherwise
the higher mode $M_{2,3}$ intercepts the threshold.
Still higher modes of the flat bunch have the higher thresholds.
The problem is considered in more detail in next section and is illustrated by Fig.~5.

Obtained thresholds of the $M_{0,1}$ and $M_{2,3}$ modes are plotted against the SC tune shift 
in Fig.4 by solid lines.
Similar curves for the square well model has been taken in Ref.~\cite{Ba2} and plotted by dashed lines.
The same results could be obtained also with help of Eq.~(18) by the substitution $U^2=4/\pi^2$.
It does not effect the bunch shape but changes characteristics of synchrotron oscillations, which looks as a secondary factor in the case.     
\begin{figure}[t!]
 \includegraphics[width=90mm]{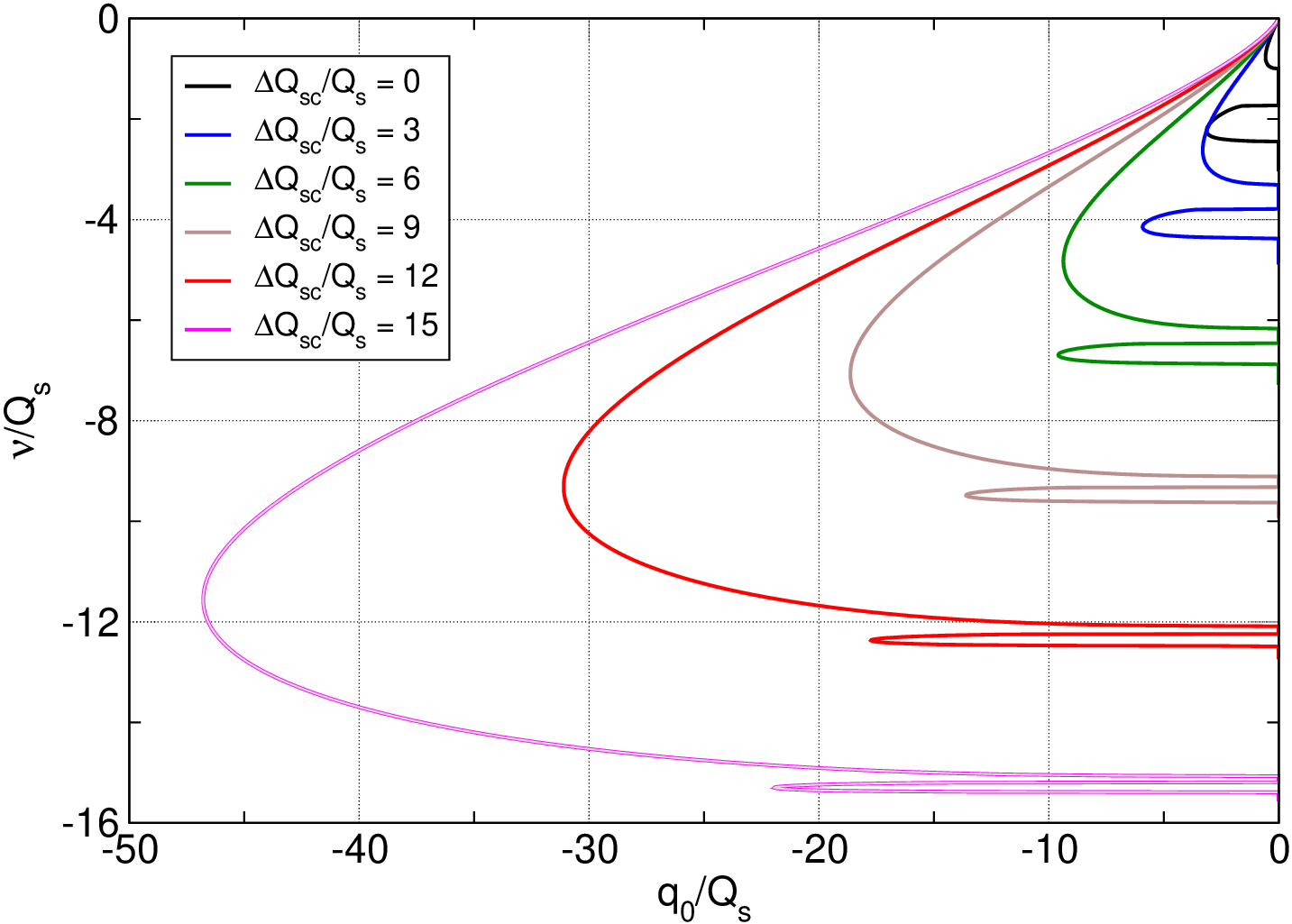}
 \caption{Tunes of the modes $M_{0,1}$ and $M_{2,3}$ against the wake strength at flat bunch with negative wake.
 Mode $M_{0,1}$ has the lower threshold at $\Delta Q_{sc}/Q_s<6$, otherwise $M_{0,1}$ threshold is less. 
 Other modes are more stable in any case.} 
 \end{figure}
 
 \begin{figure}[t!]
 \includegraphics[width=90mm]{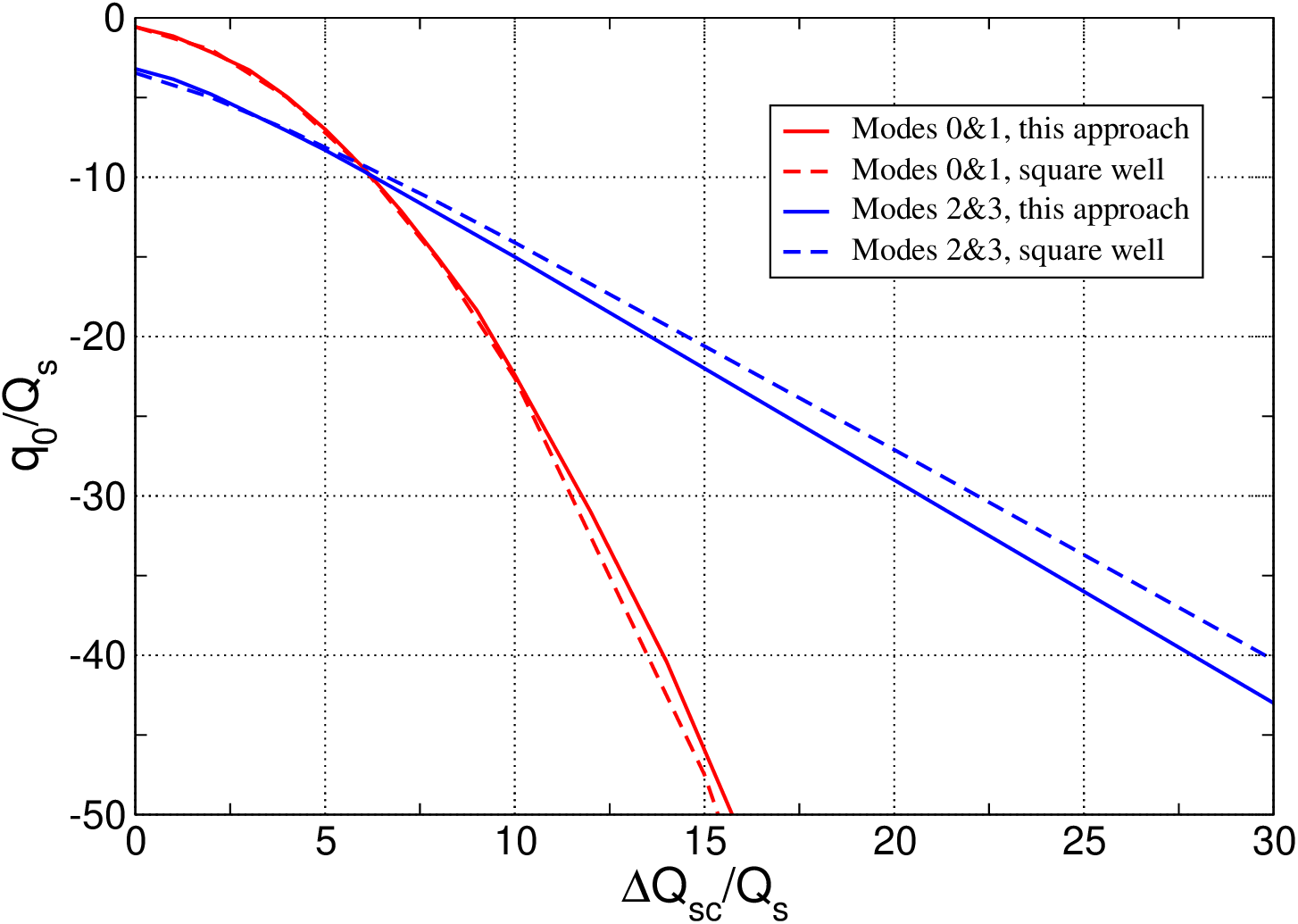}
 \caption{TMCI thresholds of the flat bunch. Solid lines are obtained in this paper, 
dashed lines -- with the square well model \cite{Ba2}. Red and blue lines represent the modes $M_{0,1}$ and $M_{2,3}$.  } 
 \end{figure}

\newpage

\section{Non-flat bunches}

The flat bunch model ignores the fact that the SC tune shift depends on longitudinal coordinate of the particle, which dependence creates an additional betatron tune spread.
It is known that the spread due to chromaticity or nonlinearity of betatron oscillations leads to appearance or a change of transverse instability thresholds.
Such an effect in accelerators is known as Landau damping which is not a dissipative process  but the absence of instability due to redistribution of the particle tunes \cite{L1}.
The issue will be considered in this section with regard to an influence of the SC tune spread on the TMCI threshold.

It follows from Fig.~3 that the lowest head-tail coherent tunes of the flat bunch join together at $\,\nu >\!\!-\Delta Q_{sc}$, that is at $\,|\nu|\!<\!\Delta Q_{sc}$, if 
$\Delta Q_{sc}/Q_s\!\!>\!\!3$.  
For a non-flat bunch, such a coupling would occur within the incoherent tune spread area.
Indeed, the particle betatron tunes $Q_\beta$ are located in the area 
$\; Q_0-\Delta Q_{max}\!\!< Q_\beta\!< Q_0\;$ where $\;\Delta Q_{max}\!=\!\Delta Q_{sc}(0)\;$ is the SC tune shift in the bunch center.
In the used notation, it means that $-\Delta Q_{max}\!\!<\nu_\beta<0$. 
Any coherent bunch oscillations, including the TMCI, are impossible in this area without an external powering, because their energy would transform at once in an incoherent form that is in the beam heating.

Actually, the space charge effect appears here as an exclusion of the bunch coherent eigentunes from the area where the particle incoherent tunes are located. 
It leads to a change of the wake field required for the eigentune coalescence, that is to the change of the TMCI threshold.
It is not a dissipative process because all the tunes are real values both before and at the coalescence.

Formally, the SC tune spread can lead to appearance of a singularity in Eq.~(8) and further, 
if $\,\nu\,$ is real number satisfying the condition $-\Delta Q_{max}<\nu<0$. 
This circumstance does not discard the method because $\nu$ appears in the equations as a parameter of the Laplace transformation.
Therefore it must have a positive imaginary part in the beginning, and  any extension to real values is allowable only as the analytical continuation of the functions.
It is seen now that the continuation is possible at either $\nu<\!\!-\Delta Q_{max}$ or $\nu>0$.
The second inequality directs us to the area where a positive wake can provide the tune coalescence and the instability. 
In particular, it means that the TMCI threshold of positive wake can't be very sensitive to the SC tune spread, that is to the bunch shape.
Therefore only the case of negative wakes is analyzed below. 

We will consider the bunch distribution functions 
\begin{eqnarray}
F = \frac{2\alpha+1}{2\pi}(1-A^2)^{\alpha-1/2}\times\bigg\{{1\quad{\mbox at}\quad A < 1 \atop 0\quad{\mbox at}\quad A\ge 1}
\end{eqnarray}
where  $A=\sqrt{x^2+v^2}$ is amplitude of synchrotron oscillations. 
Then, using Eq.~(7a) and Eq.~(12), obtain
\begin{eqnarray}
\rho(x)=C_\alpha (1-x^2)^\alpha,\qquad U^2(x) = \frac{1-x^2}{2(\alpha+1)}
\end{eqnarray}
with the normalizing coefficient
\begin{eqnarray}
C_\alpha = \frac{2\alpha+1}{2\pi}\int_{-1}^1 (1-t^2)^{\alpha-1/2}\,dt.
\end{eqnarray}
Series of Eq.~(14) with boundary conditions Eq.~(15) has been resolved by the method described at the end of Sec.~II. 
The results are represented in Fig.~5 for the cases: $\,\alpha = 0,\, C_0=1/2$ (flat bunch); $\,\alpha = 1/2,\,C_{1/2}=2/\pi$ (waterbag model); and  $\,\alpha = 1,\, C_1=3/4$ (parabolic bunch).
Eigentunes are plotted at the area $-2<\nu+\Delta Q_{max}<0$ for negative wake.

\begin{figure}[t!]
\begin{minipage}{0.32\linewidth} 
\includegraphics[width=52mm]{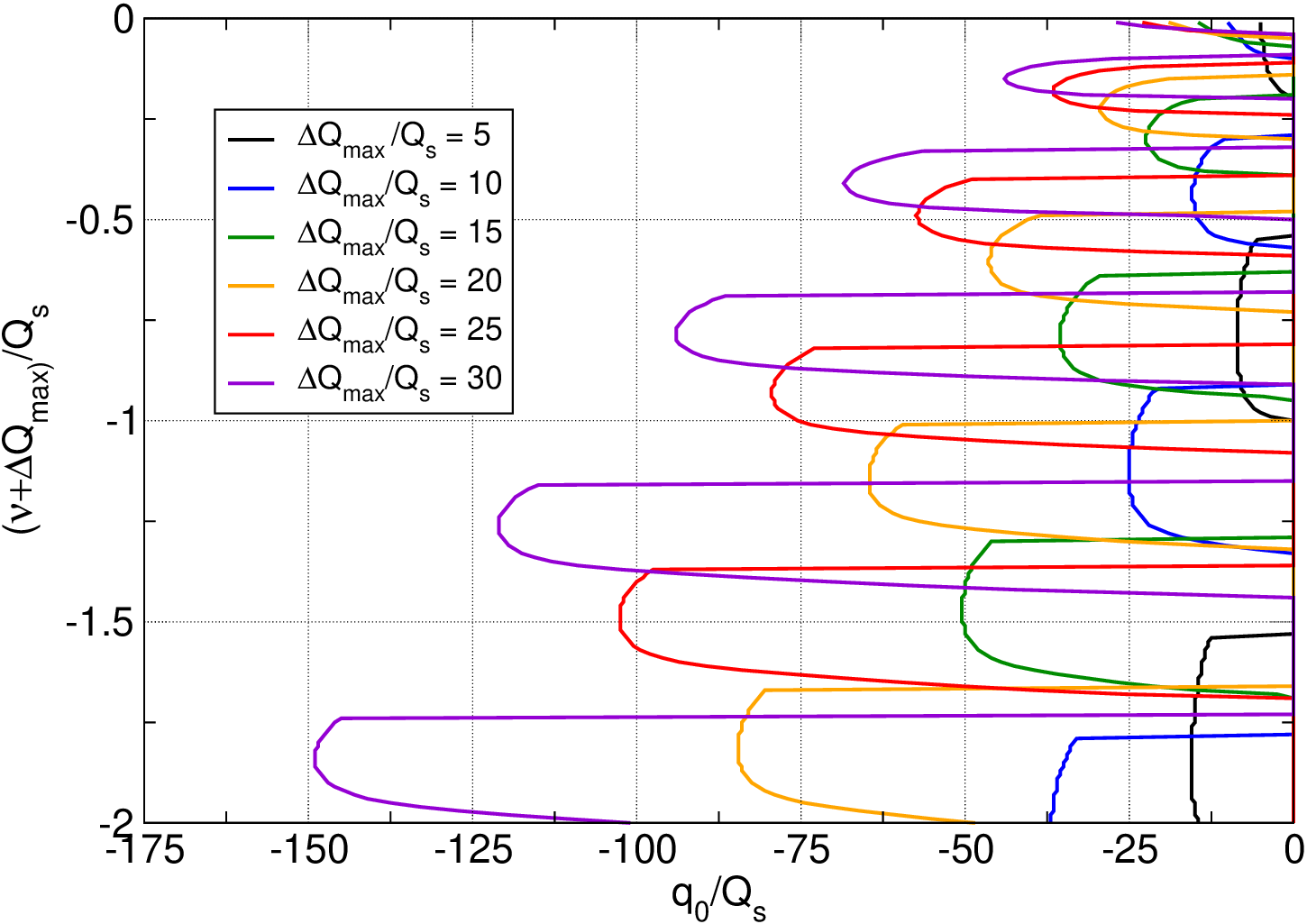}
\end{minipage}
\begin{minipage}{0.32\linewidth} 
\includegraphics[width=52mm]{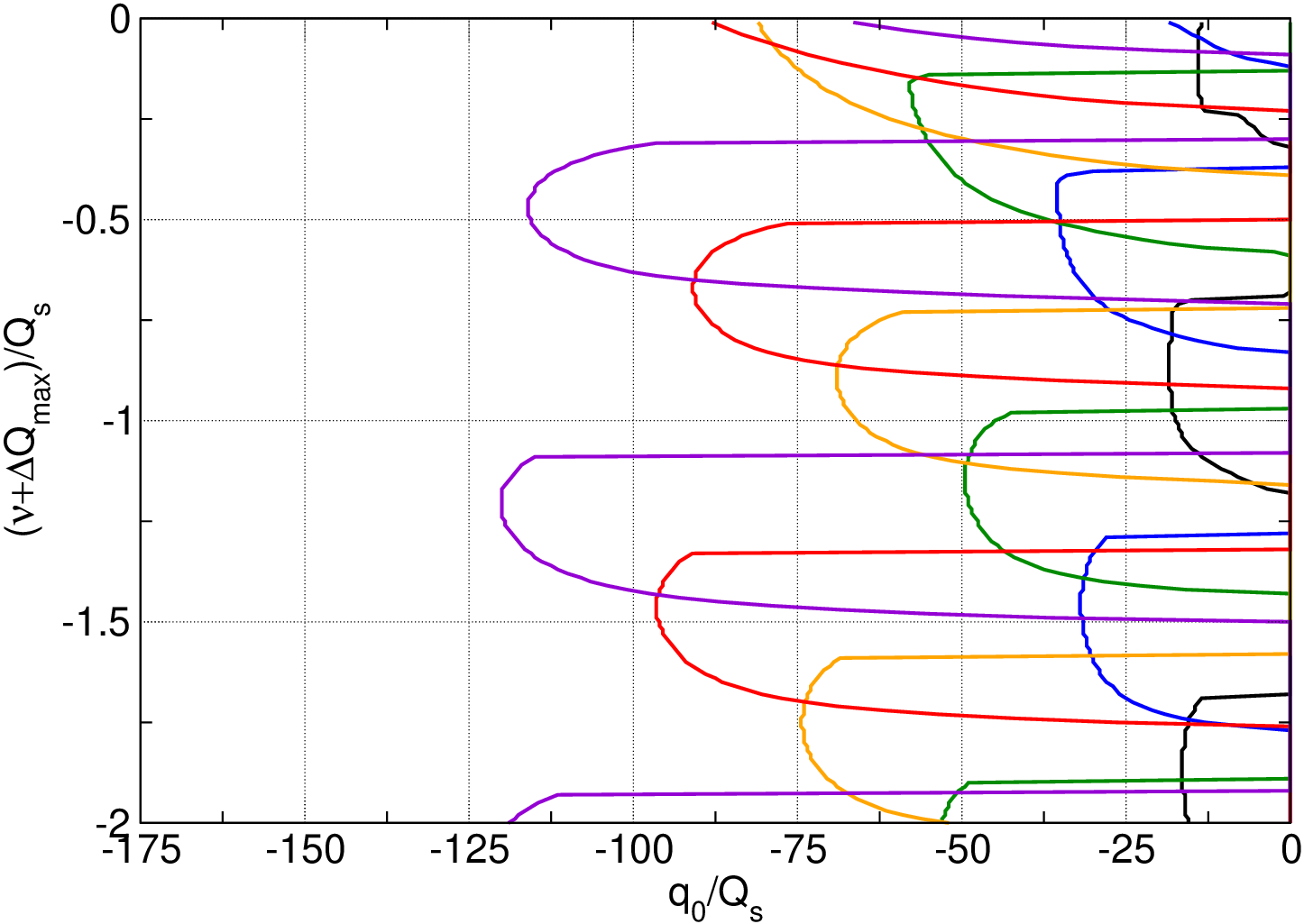}
\end{minipage}
\begin{minipage}{0.32\linewidth} 
\includegraphics[width=52mm]{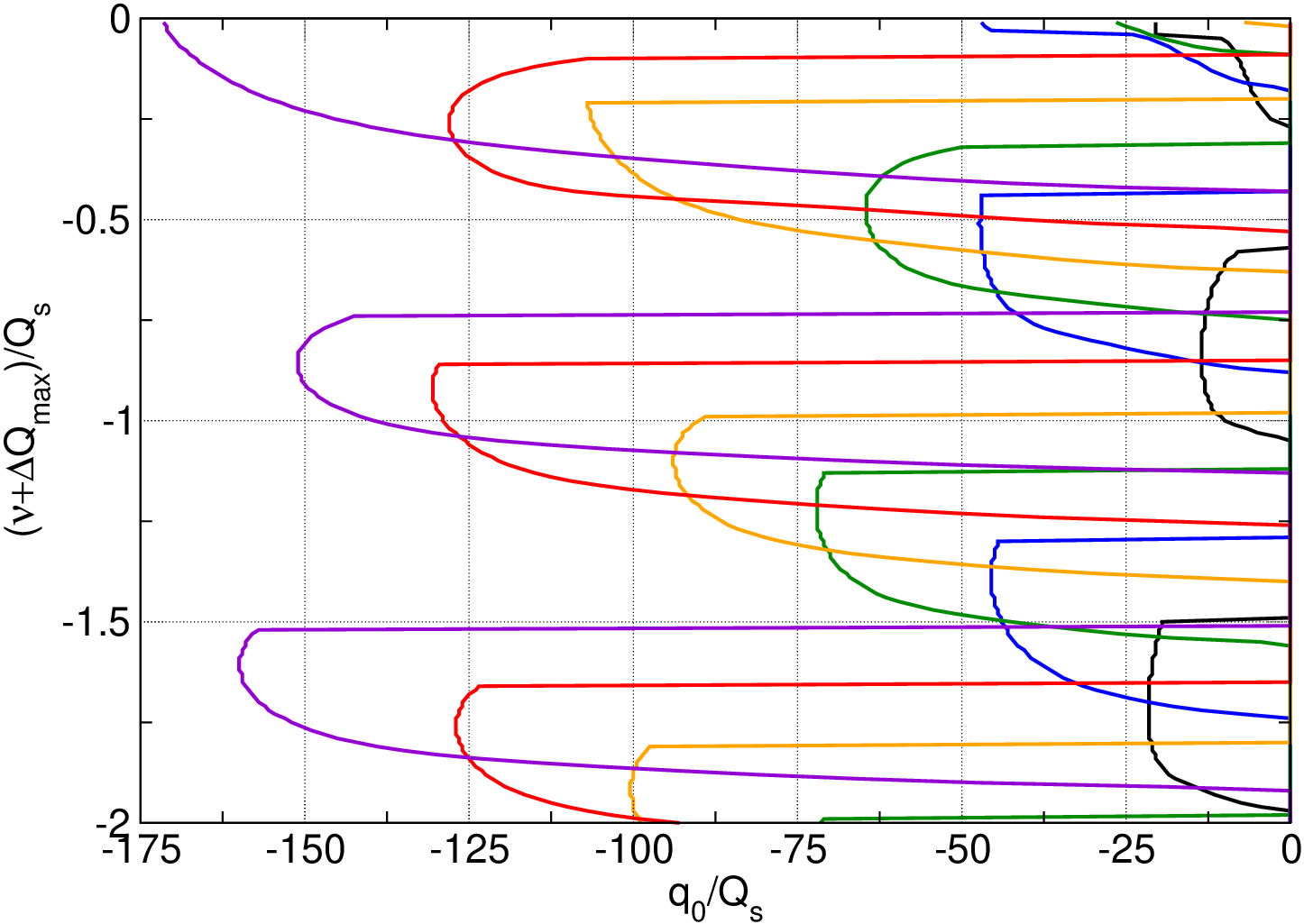}
\end{minipage}
\caption{Eigentunes of different bunches $\hat\nu_m=\nu_m+\Delta Q_{max}$ against the wake strength $q_0$ 
at different $\Delta Q_{max}$. 
Represented area $-2<\hat\nu_m<0$ is located below the particle tunes:
$\hat\nu_m<\nu_\beta$.
The top panel: flat bunch $\alpha=0$; the central panel: waterbag model $\alpha=1/2$; the bottom panel: parabolic bunch  $\alpha=1$. 
The same $\Delta Q_{max}/Q_s$ vs color are used in all the panels.} 
\end{figure}

\begin{table}[b!]
\begin{center}
\caption{Thresholds of negative wake $\,q_0/Q_s\,$ of lower TMCI modes at 
$\,\Delta Q_{max}/Q_s=15\,$ for different bunch shape.}
\vspace{5mm}
\begin{tabular}{|l|c|c|c|}
\hline 
          &~~$M_{2,3}$~~&~~$M_{4,5}~~$&~~$M_{6,7}~~$   \\   \hline
Flat      &    --22     &     --36     &     --50      \\   \hline
Waterbag  &    --57     &     --50     &     --53      \\   \hline
Parabolic &    --64     &     --73     &     --73      \\   \hline
\end{tabular}
\end{center}
\end{table}

The top panel of Fig.~5 refers again to the flat bunch. 
The upper loops of the graph represent eigentunes of the mode $M_{2,3}$ at different tune shifts.
This mode was considered also is Sec.~III where its spectrum was represented in Fig~3 by the lower loops. 
The results agree closely with each other.
For example, both plots give threshold of this TMCI mode $q_0/Q_s=-22$ at 
$\Delta Q_{sc}/Q_s=15$.
Following modes $M_{4,5}$ and $M_{6,7}$ have essentially higher in absolute value thresholds: $q_0/Q_s=-36$ and $q_0/Q_s=-50$ (the green lines in Fig.~5).

The spectra of more realistic bunch forms have a similar configuration but show much weaker dependence of the instability threshold on the mode number.
Two of them are represented also in Fig.~5.
The waterbag model $\alpha=1/2$ is represented in central panel, and the parabolic model $\alpha=1$ is illustrated by the lower picture.
Three coupled modes are shown on each of these graphs at different $\Delta Q_{max}$. 
Their thresholds are distinct from each other with no more then 10\% that is essentially less than in the flat bunch.
It is visible as well that dependence of the threshold on the mode number can be nonmonotonic.
The same follows from Table I where numerical value of the threshold is given at $\,\Delta Q_{max}/Q_s=15$.

Consideration of other distributions supports these conclusions.
The results are collected in Fig.~6 where dependence of the TMCI threshold of the most unstable mode on maximal value of the SC tune shift is plotted for distributions with different $\alpha$.
It is seen that the TMCI threshold rises when the parameter $\alpha$ increases. 
Note that increase of $\alpha$ means decrease of the bunch r.m.s length. 
Really, it follows from Eq.~(20) 
$$
\sigma^2=\frac{2\alpha+1}{2\pi}\int\limits_0^1 (1-A^2)^{\alpha+1/2}\!A^3dA
\int\limits_{-\pi}^\pi\!\cos^2\!\phi d\phi=\frac{1}{2\alpha+3}
$$
It means that, at the same central density, the shorter bunch have the higher TMCI threshold, and that the flat bunch model provides the lower estimation of the threshold. 
\\

\begin{figure}[b!]
\includegraphics[width=100mm]{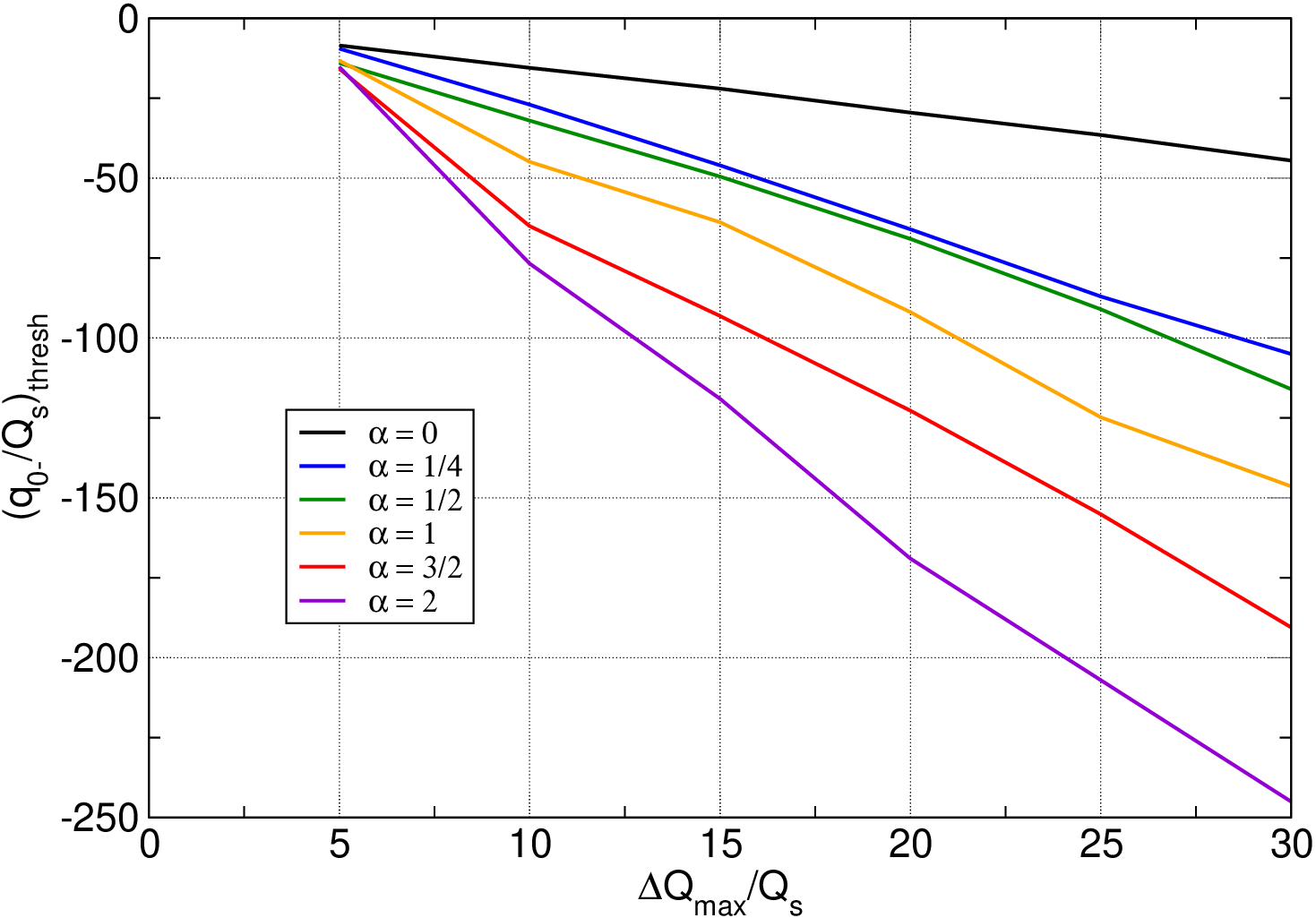}
\vspace{-5mm}
\caption{TMCI thresholds of different form bunches against the SC tune shift.} 
\end{figure}

\section{Conclusion}

It is shown in the paper that the TMCI thresholds of the flat bunch only slightly depends on the potential well shape being almost identical in the square well and in the parabolic one,
if the same ratio of the space charge tune shift to the synchrotron tune $\Delta Q_{sc}/Q_s$ is used.
It means that the characteristic of synchrotron oscillations is the secondary factor in the case.  
The space charge tune shift reduces the TMCI threshold of positive wakes, and increases it if the wake is negative.
In the last case, the most unstable coupled modes of the flat bunch are: the lowest mode $M_{0,1}$ at $\Delta Q_{sc}/Q_s<6$, and the next mode $M_{2,3}$ at $\Delta Q_{sc}/Q_s>6$. 
Corresponding tunes are located a little higher or lower of the betatron tune of the particles, with the SC shift included.  
All other modes have higher threshold at any value of the shift.

Spread of the SC tune shift, intrinsic to any real (non-flat) bunch, essentially changes the bunch spectrum at negative wake field.
It forces out the coherent mode tunes from the area $-\Delta Q_{max}\!\!<\!\nu\!<\!0$ where the particle incoherent tunes are located. 
As a result, the mode $\,M_{0,1}\,$ is excluded from the spectrum at $\,\Delta Q_{max}>Q_s$, 
and tunes of other coherent modes fall below the minimal particle tune. 

In such conditions, thresholds of all the TMCI modes differ from each other to only a small extension.
These thresholds increase when r.m.s. length of the bunch decreases at fixed total length.
Therefore the flat bunch model with given $\Delta Q_{max}$ provides only the lowest estimation of the TMCI threshold.

\section{Acknowledgment}

Fermi National Accelerator Laboratory is operated by Fermi Research Alliance,
LLC under Contract No. DEAC02-07CH11395 with the United States Department of Energy.

\end{document}